\documentclass[pra,superscriptaddress,longbibliography,twocolumn]{revtex4-2}
\usepackage[utf8]{inputenc}
\usepackage[english]{babel}
\usepackage{amsmath}
\usepackage{amsfonts}
\usepackage{amssymb}
\usepackage{graphicx}
\usepackage{amsthm}
\usepackage{color}
\usepackage{hyperref}
\usepackage[caption=false]{subfig}
\usepackage{graphicx,float}% Include figure files
\usepackage{dcolumn}% Align table columns on decimal point
\usepackage{bm}% bold math
\usepackage{mathrsfs}
\usepackage{txfonts}
\usepackage{CJK}
\usepackage[amssymb]{SIunits}
\usepackage{algorithm}
\usepackage{algpseudocode}
\usepackage{epsfig}
\usepackage{epstopdf}
\usepackage{lipsum}
\usepackage{placeins}
\usepackage{physics}
\usepackage{chngcntr}

\newenvironment{SChinese}{%
\CJKfamily{gbsn}%
\CJKtilde
\CJKnospace}{}
\allowdisplaybreaks[2]

\begin{document}

\begin{CJK}{UTF8}{}
\begin{SChinese}

\title{Photonic indistinguishability characterization and optimization for cavity-based single-photon source}

\author{Miao Cai}  %
 \affiliation{College of Engineering and Applied Sciences,  Nanjing University, Nanjing 210023, China}
 \affiliation{National Laboratory of Solid State Microstructures,  Nanjing University, Nanjing 210093, China}
 
\author{Mingyuan Chen}  %
 \affiliation{College of Engineering and Applied Sciences,  Nanjing University, Nanjing 210023, China}
 
\author{Jiangshan Tang}  %
 \affiliation{College of Engineering and Applied Sciences,  Nanjing University, Nanjing 210023, China}

\author{Keyu Xia (夏可宇)}  %
 \email{keyu.xia@nju.edu.cn}
 \affiliation{College of Engineering and Applied Sciences,  Nanjing University, Nanjing 210023, China}
 \affiliation{National Laboratory of Solid State Microstructures,  Nanjing University, Nanjing 210093, China}
 \affiliation{Shishan Laboratory, Suzhou Campus of Nanjing University, Suzhou 215000, China}

\begin{abstract}
Indistinguishability of single photons from independent sources is critically important for scalable quantum technologies. We provide a comprehensive comparison of single-photon indistinguishability of different kinds of cavity quantum electrodynamics (CQED) systems by numerically simulating Hong-Ou-Mandel (HOM) two-photon interference. We find that the CQED system using nature atoms exhibit superiority in indistinguishability, benefiting from the inherently identical features. Moreover, a $\Lambda-$type three-level atoms show essential robust against variation of various system parameters because it exploits the two ground states with considerable smaller decay rates for single-photon generation.  Furthermore, a machine learning-based framework is proposed to significantly and robustly improve single-photon indistinguishability for non-identical two CQED systems. This work may pave the way for designing and engineering reliable and scalable photon-based quantum technologies.
\end{abstract}

\maketitle

\end{SChinese}
\end{CJK}
%%%%%%%%%%%%%%%%%%%%%%%%%%%%%%%%%%%%%%%%%%%%%%
\section{Introduction}\label{sec:intro}
Single-photon source serves as a fundamental cornerstone in numerous quantum technologies, including optical quantum computing~\cite{Jeremy2007,Kok2007,Zhong2020,Jinpeng2021}, boson sampling~\cite{Marco2015}, quantum cryptography protocols~\cite{Beveratos2002,Alleaume2004,Takesue2007,Bozzio2022}, and quantum communication applications~\cite{Yuan2010,Hu2016,Couteau2023}. The development and performance of photon-based quantum information technology depend on single-photon sources with high-quality attributes, including a high emission efficiency, a perfect single-photon purity, and photon indistinguishability~\cite{Michler2000,Kuhn2010,Aharonovich2016,Lu2021}.

Extensive efforts have been made to generate single photons within various systems, such as single trapped atoms~\cite{Darque2005,Julian2012}, ions~\cite{Moehring2007,Gerber2009}, single molecules~\cite{Brunel1999,Lounis2000}, nonlinear wave mixing~\cite{Yuan2011,Silverstone2014}, and cavity quantum electrodynamics (CQED) systems~\cite{Kuhn2002,McKeever2004,Kuhn2016}. Among these systems, CQED systems using quantum dots and three-level atoms have achieved significant success in generating single photons~\cite{He2017,He2019,Wang2019,McKeever2004,Barak2008,Aoki2009}. By coupling an atom system to optical cavity, the CQED system can deterministically generate single photons and greatly enhance single-photon emission efficiency and single-photon purity~\cite{Purcell1995,Kaupp2016,Somaschi2016,Benedikter2017}. However, the inherent randomness in the self-assembled growth process of artificial atoms causes fluctuations in the CQED system's properties such as resonance frequency and cavity-atom coupling strength. This fabrication imperfection results in degradation of single-photon indistinguishability and limits the scalability of quantum information technologies based on deterministic single-photon generation~\cite{Eisaman2011,Reindl2017}. Therefore, production of identical artificial atoms is highly desirable, but challenging. A recent solution enhancing indistinguishability of photons is to find similar artificial atoms from a large number of atoms~\cite{Zhai2022}. Thus, it is crucially important to understand the influence of the CQED system structures and parameters on the indistinguishability of single photons in the current intense competition for sources.

For solid-state single-photon sources, it is challenging to flexibly adjust system parameters once they are fabricated. Therefore, finding the optimal system parameters is of great importance, but it has been less explored so far. Furthermore, a flexible approach for optimizing photon properties is highly valuable, but still poses a challenge. In particular, the CQED system using three-level atoms can tailor the temporal shape of the emitted single-photon wavefunction. By appropriately controlling the driving field, a single photon with a corresponding wavefunction is emitted from the system via stimulated Raman adiabatic passage (STIRAP)~\cite{Kuhn2002,McKeever2004,Kuhn2010}. This feature enables the optimization of single-photon indistinguishability by adjusting the driving field without the need to re-manufacture the CQED system. Machine learning (ML), which has become a powerful tool for designing, optimizing, and controlling quantum systems~\cite{Biamonte2017,Bukov2018,Niu2019,Sivak2022,Alexey2018,Krenn2020,Cai2021,Cai2022}, offering a general and flexible approach to tackling this challenging task.

In this work, we simulate Hong-Ou-Mandel (HOM) interference between single photons emitted from CQED systems to investigate the impact of cavity and atom parameters on single-photon indistinguishability. Our results indicate the single-photon indistinguishability superiority of the CQED system using natural atoms. We also demonstrate that a CQED system using three-level atoms exhibits greater robustness to atom decay and shows near-perfect single-photon indistinguishability across a wider range of system parameters compared to a CQED system with two-level atoms. Moreover, we propose a ML framework to optimize the single-photon indistinguishability for CQED system with a three-level atom by finding the optimal driving field. We also validate the effectiveness of this framework for CQED systems with parameter fluctuations. Our work provides insights into the performance and potential applications of CQED-based single-photon sources, and opens up a new route to engineering flexible and reliable quantum networks in an ``intelligent'' way.

This paper is organized as follows: In Sec.~\ref{sec:SysModel}, we first present the theoretical models of single-photon emission schemes in the CQED system using two-level and $\Lambda$-type three-level atoms, and simulation method of Hong-Ou-Mandel two-photon interference. In Sec.~\ref{sec:MLframe}, we demonstrate our ML framework for enhancing the photon indistinguishability of CQED-based single-photon sources. In Sec.~\ref{sec:Analysis}, we provide a comprehensive analysis of the influence of various system parameters on single-photon properties within two-level and  $\Lambda$-type CQED systems. In Sec.~\ref{sec:MLresult}, we show the improvement of indistinguishability of single photons with our ML framework. Finally, we conclude with a discussion in Sec.~\ref{sec:discussion}.

%%%%%%%%%%%%%%%%%%%%%%%%%%%%%%%%%%%%%%%%%%%%%%%%%%%%%%%%%%%%%%
\section{CQED-based single-photon source}\label{sec:SysModel}
Among CQED systems, atoms (both artificial and natural) with two-level and three-level structures are widely utilized to generate single photons through coupling with optical cavities. Below, we first introduce a single-photon generation scheme in a two-level CQED system. Next, we consider a more complex case in which the atom has a $\Lambda$-type three-level structure in the system, namely $\Lambda$-type CQED system. After that, we elaborate on the assessment method of single-photon indistinguishability by HOM interference.

\subsection{Two-level CQED system}\label{sec:Sys1}
\begin{figure}
  \centering
  \includegraphics[width=1.0\linewidth]{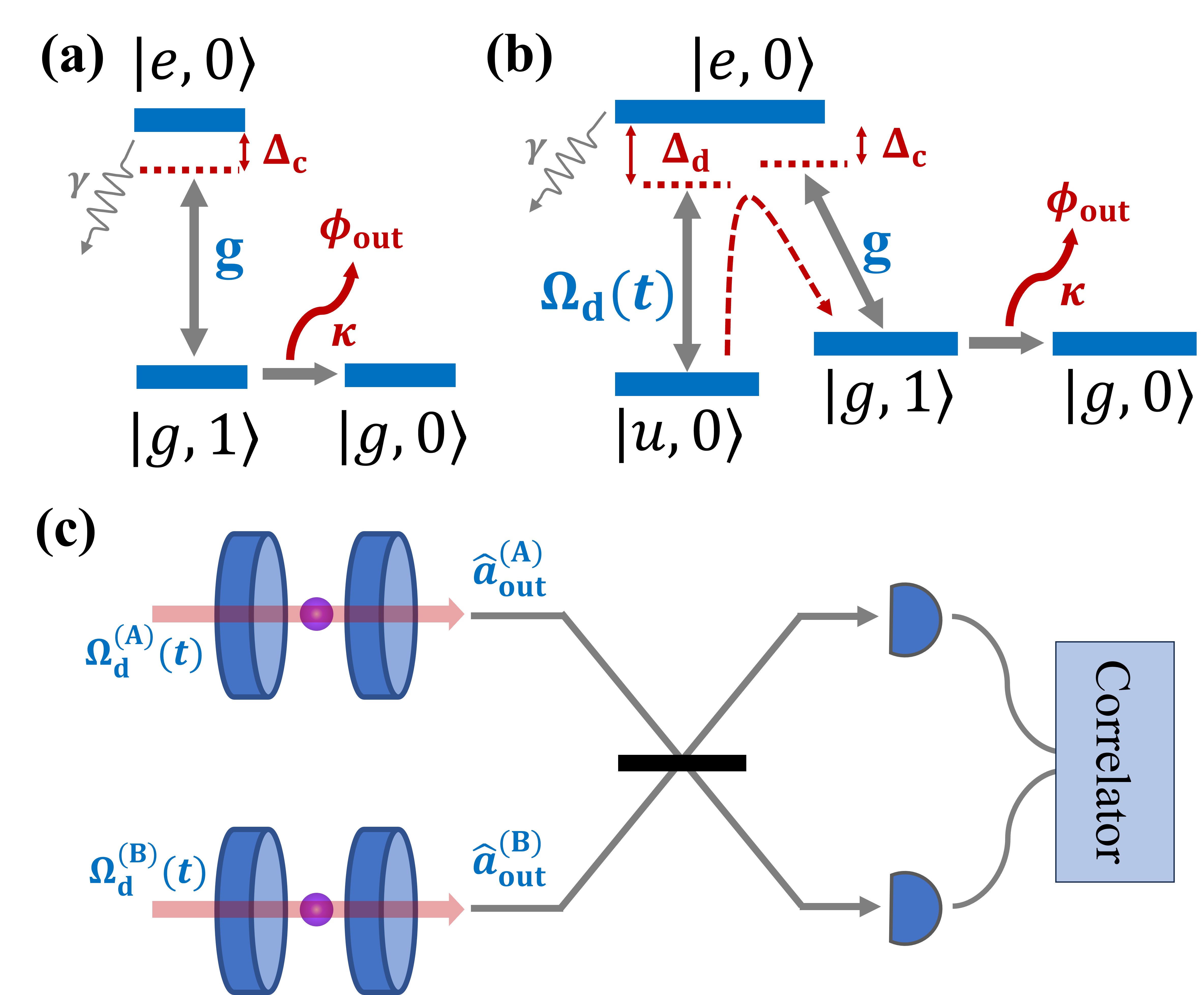} \\
  \caption{Single-photon generation schemes in (a) two-level CQED system and (b) $\Lambda$-type CQED system. (c) Schematic of the HOM interference. The emissions from two independent single-photon sources are interfered on two detectors by a beamsplitter.}
  \label{fig:FIG1}
  \end{figure}

The schematic of the two-level CQED system is depicted in Fig.~\ref{fig:FIG1}(a). An atom is trapped in a single-mode optical cavity with frequency $\omega_{\text{c}}$. The atom has a ground state $\ket{g}$ and an excited state $\ket{e}$ with atomic resonance frequencies $\omega_{\text{g}}$ and $\omega_{\text{e}}$, respectively. The states $\ket{0}$ and $\ket{1}$ denote a cavity field with zero and one photon, respectively. Under the rotating frame transformation, the Hamiltonian of the system can be expressed as,
\begin{equation}
\label{eq:Eq1}
\begin{aligned}
H = \Delta_{\text{c}}\hat{a}^{+}\hat{a} + g(\hat{a}^{+}\sigma + \sigma^{+}\hat{a}) \; ,
\end{aligned}
\end{equation}
where $\hat{a}^{+}$, $\hat{a}$, $\sigma^{+}$, $\sigma$ represent the creation and annihilation operators of cavity modes and atomic excitation, respectively. $\Delta_{\text{c}}$ defined by $\Delta_{\text{c}} = \omega_{\text{e}}-\omega_{\text{g}}-\omega_{\text{c}}$ is the frequency detuning between the atom and the cavity.

The dynamics of the system is determined by the Lindblad master equation, which can be written as,
\begin{equation}
\label{eq:Eq2}
\begin{aligned}
\dot{\rho}(t) =& -\frac{i}{\hbar}\left[H,\rho\right] \\
&+ \frac{1}{2}\sum_{n}\left[2\hat{C}_{n}\rho(t)\hat{C}_{n}^{+} - \rho(t)\hat{C}_{n}^{+}\hat{C}_{n} - \hat{C}_{n}^{+}\hat{C}_{n}\rho(t)\right] \; ,
\end{aligned}
\end{equation}
where $C_{n} = \{\sqrt{\kappa}\hat{a},\sqrt{\gamma}\sigma\}$ are the collapse operators. Here $\kappa$ is the decay rate via the optical cavity and $\gamma$ is the collapse rate corresponding to spontaneous emission of the atom. To achieve single-photon emission, a sudden excitation process such as a short $\pi$ pulse with large Rabi frequency firstly drives the system into its excited state. Subsequently, a photon gets spontaneously emitted into the cavity, and then transmit out through the cavity decay channel. Thus the output field operator can be defined as $\hat{a}_{\text{out}} = \sqrt{\kappa}\hat{a}$ and the emitted photon wavefunction is $\phi_{\text{out}}(t) = \langle \hat{a}_{\text{out}}^{\dag}(t) \hat{a}_{\text{out}}(t) \rangle$.

\subsection{$\Lambda$-type CQED system}\label{sec:Sys2}
The two-level CQED system has the simplest atom structure. To advance the comprehensive study of CQED-based single-photon sources, we extend our investigation to the $\Lambda$-type CQED system, in which the atom has a $\Lambda$-type three-level structure. The schematic of single-photon generation in a $\Lambda$-type CQED system is depicted in Fig.~\ref{fig:FIG1}(b). The atom has two long-lived ground states, denoted as $\ket{u}$ and $\ket{g}$ and an excited state $\ket{e}$. The cavity mode is near resonant with the atomic transition between states $\ket{e}$ and  $\ket{g}$, but far off resonance from the transition between $\ket{u}$ and  $\ket{g}$. Thus, the product states $\ket{e,0}$ and $\ket{g,1}$ are coupled by the cavity mode with a coupling strength of $g$. Here, the coupling strength $g$ is considered constant. 

To generate a single photon, the atom is exposed to a time-dependent classic laser field $\Omega_{\text{d}}(t)$ (referred to as the driving field) with frequency $\omega_{\text{d}}$. The driving field is near resonant with the atomic transition between the states $\ket{u}$ and $\ket{e}$, thereby coupling the product states $\ket{u,0}$ and $\ket{e,1}$. Driven by this field, the system transitions from the initial state $\ket{u,0}$ to the excited state $\ket{e,1}$. Then, the quantum excitation is transferred to the cavity mode $\ket{g,1}$ through the atom-cavity coupling. At last, the photon is emitted into the output channel through cavity decay, generating a single photon with a temporal wavefunction $\phi_{\text{out}}(t)$ and leaving the system in the state $\ket{g,0}$.

For convenience, we use subscripts $\{1,2,3\}$ to relate parameters and variables to the basis $\{\ket{u,0},\ket{e,0},\ket{g,1}\}$. The original Hamiltonian of the combined CQED system can then be expressed as follows,
\begin{equation}
\label{eq:Eq3}
\begin{aligned}
&H = \omega_{\text{c}}\hat{a}^{+}\hat{a} + \sum_{i=1}^{3}\omega_{i}\sigma_{ii} + \Omega_{\text{d}}(t)\left(e^{-i\omega_{\text{d}}t}\sigma_{21} + e^{i\omega_{\text{d}}t}\sigma_{12}\right)\\
&\qquad+ g\left(\hat{a}^{+}\sigma_{32} + \hat{a}\sigma_{23}\right) \; ,
\end{aligned}
\end{equation}
where $\omega_{\text{c}}$ and $\omega_{\text{d}}$ are the frequencies of the cavity mode and driving field, respectively. The atomic resonance frequencies of the three states $\{\ket{u,0},\ket{e,0},\ket{g,1}\}$ are denoted as $\omega_{i}$, where $i=1,2,3$. Under the rotating frame transformation with the rotation operator $U = \exp{i\left(\omega_{2}-\omega_{3}\right)\hat{a}^{+}\hat{a} + i\sum_{i=1}^{3}\omega_{i}\sigma_{ii}}$, the Hamiltonian can be rewritten as,
\begin{equation}
\label{eq:Eq4}
\begin{aligned}
&H = \Delta_{\text{c}}\hat{a}^{+}\hat{a} \\
&\qquad+ \Omega_{\text{d}}(e^{-i\Delta_{\text{d}}t}\sigma_{21} + e^{i\Delta_{\text{d}}t}\sigma_{12}) + g(\hat{a}^{+}\sigma_{32}+\hat{a}\sigma_{23}) \; ,
\end{aligned}
\end{equation}
where $\Delta_{\text{c}}$ is the detuning between the $\ket{e,0}\leftrightarrow\ket{g,1}$ transition and the coupling cavity, and $\Delta_{\text{d}}$ represents the detuning between the $\ket{u,0}\leftrightarrow\ket{e,0}$ transition and the driving field, as illustrated in Fig.~\ref{fig:FIG1}(b).

The process of single-photon generation is also determined by the Lindblad master equation, as described in Eq.~\ref{eq:Eq2}. The collapse operators in the $\Lambda$-type CQED system are given by $C_{n} = \{\sqrt{\kappa}\hat{a},\sqrt{\gamma_{12}}\sigma_{12},\sqrt{\gamma_{32}}\sigma_{32}\}$. Here, $\kappa$ represents the decay rate through the optical cavity, while $\gamma_{12}$ and $\gamma_{32}$ denote the collapse rates associated with spontaneous emission from the excited state to the system's two ground states. Therefore, the overall rate of collapse $\gamma$ of the excited state is given by $\gamma = \gamma_{12} + \gamma_{32}$. In the following investigation, we assume that atom decay rates satisfy fixed proportional relationships, where $\gamma_{12} = \frac{5}{9}\gamma$ and $\gamma_{32} = \frac{4}{9}\gamma$. In the $\Lambda$-type CQED system, a single photon is emitted through the cavity decay channel, similar to the process in the two-level CQED system. Thus, the definition of the output field operator $\hat{a}_{\text{out}}$ and the single-photon wavefunction $\phi_{\text{out}}(t)$ are the same as in Sec.~\ref{sec:Sys1}.

\subsection{HOM two-photon interference}\label{sec:Sys3}
HOM two-photon interference is one of the most widely used standard tools to investigate the indistinguishability between photons. In a HOM interferometer, two independent photon sources are interfered on two detectors by using a single $50-50$ beamsplitter. The interference visibility reaches unity when the photon inputs are perfectly identical, and decreases as the inconsistency in photon properties increases. Thus, the indistinguishability of two CQED-based single-photon sources can be investigated by simulating the HOM interference visibility. As shown in Fig.~\ref{fig:FIG1}(c), two CQED systems (denoted by notations $(A)$ and $(B)$) are periodically driven by the driving fields $\Omega_{\text{d}}^{(\text{A})}(t)$ and $\Omega_{\text{d}}^{(\text{B})}(t)$, and their outputs $\hat{a}_{\text{out}}^{\text{(A)}}(t)$,$\hat{a}_{\text{out}}^{\text{(B)}}(t)$, incident into the HOM interferometer. In the case of the two-level CQED system, it is assumed that the excitation pulse is short enough for its temporal shape to be irrelevant. Therefore, we set the initial state of the two-level CQED system as the excited state in the following simulation for convenience.

\begin{figure}
  \centering
  \includegraphics[width=1.0\linewidth]{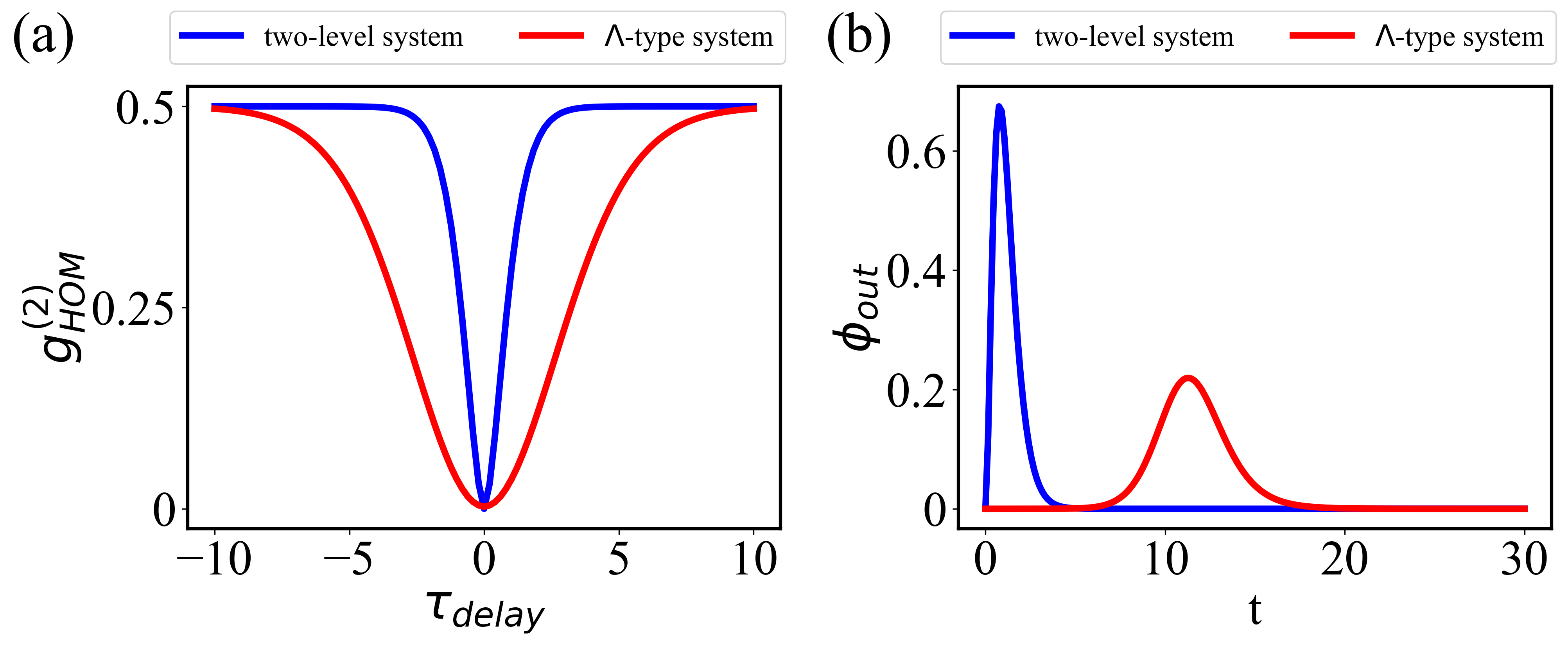} \\
  \caption{(a) Simulated HOM correlation $g^{(2)}_{\text{HOM}}$ is plotted against the delay time $\tau$ for perfectly identical single photons emitted from two-level (blue) and $\Lambda$-type (red) CQED systems. (b) Typical single-photon wavefunctions generated by a two-level (blue) and a $\Lambda$-type (red) CQED system.}
  \label{fig:FIG2}
  \end{figure}

In the numerical simulation, the visibility of the HOM interference, denoted as $V$, is defined by $V = 1- g^{(2)}_{\text{HOM}}[0]/\underset{\tau\rightarrow T}{\text{lim}}g^{(2)}_{\text{HOM}}[\tau]$. Here, $g^{(2)}_{\text{HOM}}[\tau]$ represents the normalized correlation of two independent single-photon sources with a delay time of $\tau$. It can be obtained by calculating the following equation~\cite{Fischer2016},
\begin{equation}
\label{eq:Eq5}
\begin{aligned}
& g^{(2)}_{\text{HOM}}[\tau] = \\
&\quad\frac{1}{2}\left(1 - \text{Re}\int_{0}^{T}\int_{0}^{T}dtdt^{\prime}\left[G_{\text{(A)}}^{(1)}(t,t^{\prime})\right]^{*}\left[G_{\text{(B)}}^{(1)}(t-\tau,t^{\prime})\right]\right) \; ,
\end{aligned}
\end{equation}
where $T$ is the duration time of the complete single-photon generation process. $G_{\text{(A)}}^{(1)}(t,t^{\prime}) = \langle \hat{a}_{\text{out}}^{\text{(A)}\dag}(t) \hat{a}_{\text{out}}^{\text{(A)}}(t^{\prime})\rangle$ and $G_{\text{(B)}}^{(1)}(t,t^{\prime}) = \langle \hat{a}_{\text{out}}^{\text{(B)}\dag}(t) \hat{a}_{\text{out}}^{\text{(B)}}(t^{\prime})\rangle$ represent the first-order coherences of the single-photon sources. The numerically simulated $g^{(2)}_{\text{HOM}}[\tau]$ for nearly perfect identical single photons emitted from two-level and $\Lambda$-type CQED systems are depicted in Fig.~\ref{fig:FIG2}(a). We also present typical wavefunctions of single photons generated by two types of CQED systems in Fig.~\ref{fig:FIG2}(b), where we apply a Gaussian-type driving field to the $\lambda$-type CQED system.

The expression in Eq.~\ref{eq:Eq5} can be interpreted as quantifying the overlap of the single-photon wavefunctions $\phi_{\text{out}}$ from the sources. For a $\Lambda$-type CQED system, the emitted single-photon wavefunction is determined by the driving field when the other system parameters are fixed. From Eq.~\ref{eq:Eq5} we can see that as the overlap of the two single-photon wavefunctions increases, the value of the zero-delay normalized correlation $g^{(2)}_{\text{HOM}}[0]$ decreases, leading to an increase in the interference visibility $V$. Thus, we can improve the indistinguishability of photons by adjusting the driving fields to increase the overlap between single-photon wavefunctions for $\Lambda$-type CQED-based single-photon sources.

\section{Machine learning framework}\label{sec:MLframe}
\begin{figure}
  \centering
  \includegraphics[width=1.0\linewidth]{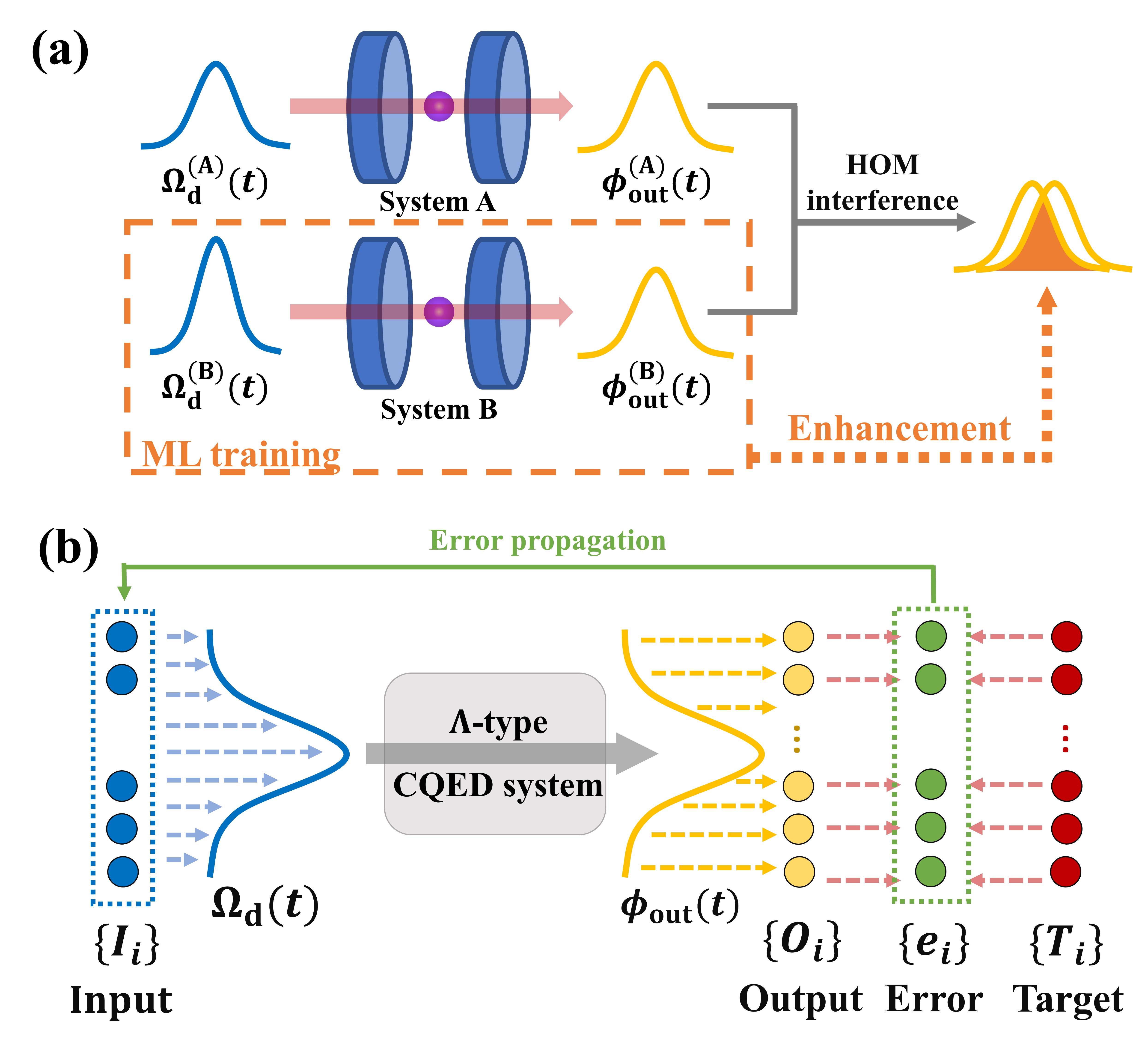} \\
  \caption{(a) ML framework working mechanism. With ML algorithm, the optimal driving field of one single-photon source is determined, which maximizes the overlap between the emitted single photon wavefunctions. This enhances the indistinguishability of photons from two single-photon sources. (b) Detailed schematic of our ML algorithm. The input dataset (driving field) iteratively adjusts itself based on feedback from the error set, which is calculated from the output and target datasets. Here, different colors represent various stages in our algorithm.}
  \label{fig:FIG3}
  \end{figure}

As we mentioned in Sec.~\ref{sec:Sys3}, the photon indistinguishability can be enhanced by increasing the overlap between single-photon wavefunctions. To accomplish this, we have developed a ML framework that can find the driving fields required to generate identical single-photon wavefunctions in non-identical $\Lambda$-type CQED systems. The operational mechanism of our ML framework is illustrated in Fig.~\ref{fig:FIG3}(a): Given a single-photon source A which emits single photons with wavefunctions $\phi_{\text{out}}^{\text{(A)}}(t)$, our ML framework is used to determine the optimal driving field for another source B to generate single photons with wavefunctions that are nearly identical, thus improving the indistinguishability of photons of the two sources.

In our framework, we utilize a reinforcement learning (RL) strategy to optimize the driving field. RL is an interdisciplinary field that combines ML and optimal control. It has recently been utilized to discover optimal strategies for various quantum technologies, including quantum error correction~\cite{Fosel2018}, quantum control~\cite{Mavadia2017,Niu2019} and quantum transport~\cite{Porotti2019}. Generally, a RL algorithm consists of three main parts: \emph{action policy}, \emph{reward signal} and \emph{value function}. In RL, an intelligent agent learns to take actions in a dynamic system and updates its action policy according to a reward signal, ultimately maximizing the value function~\cite{Sutton2018}. Here, the action policy can be understood as a sequence of actions over discrete steps. The reward signal represents the benefit or loss from an action at a given time, and the value function provides an overall evaluation of the action sequence.

In the RL training process of our framework, the action policy represents the value of the driving field at a specific time, while the reward signal corresponds to the error between the emitted and the target single-photon wavefunction in each time segment. Here, we use the HOM interference visibility $V$ between single photons emitted from systems A and B as the value function to assess the quality of the driving field.

The detailed training process in our ML framework is shown in Fig.~\ref{fig:FIG3}(b). We first initialize the discrete input dataset $\{I_{i}\}$, and then convert it into a continuous driving field function $\Omega_{\text{d}}(t)$ using linear interpolation,
\begin{equation}
\label{eq:Eq6}
\Omega_{\text{d}}(t)  = I_{i} + \frac{I_{i+1}-I_{i}}{\Delta t}(t-t_{i}), t\in [t_{i},t_{i+1}) \; ,
\end{equation}
where $\Delta t = t_{i} - t_{i-1}(i\in\{1,2,...,N\})$ is the time interval of the $i$-th time segment. With $\Omega_{\text{d}}(t)$ we can obtain the emitted single-photon wavefunction $\phi_{\text{out}}(t)$ through numerical simulation of the CQED system. Next, we convert $\phi_{\text{out}}(t)$ into a discrete output dataset $\{O_{i}\}$ using the following equation,
\begin{equation}
\label{eq:Eq7}
O_{i} = \frac{1}{\Delta t}\int_{t_{i}}^{t_{i+1}}\phi_{\text{out}}(t)dt \; ,
\end{equation}
With Eq.~\ref{eq:Eq6} and Eq.~\ref{eq:Eq7}, We establish a mapping from the input dataset to the output dataset. We then calculate the error set $\{e_{i}\}$ from the target dataset $\{T_{i}\}$ and the output dataset $\{O_{i}\}$ using the formula $e_{i} = T_{i} - O_{i}$. The error set is fed back to adjust the input dataset according to the following feedback rule,
\begin{equation}
\label{eq:Eq8}
I_{i} \leftarrow I_{i} + \eta e_{i} \; ,
\end{equation}
where $\eta\in(0,1]$ is the learning rate.

For each training step, the training process continues until the error is negligible, or the number of iterations exceeds a preset value to avoid an endless loop, or the value function $V$ decreases during training. Finally, once the training for all steps is completed, the optimized driving field is obtained. Then, by applying this driving field to the $\Lambda$-type CQED system, we can generate a single photon with a wavefunction near identical to the other system, thereby enhancing the indistinguishability of photons from two single-photon sources. By applying the above process to multiple pairs of single-photon sources, our framework can be easily extended to the case of a single-photon source network.

\section{Characterization of photon indistinguishability}\label{sec:Analysis}

\begin{figure*}
  \centering
  \includegraphics[width=1\linewidth]{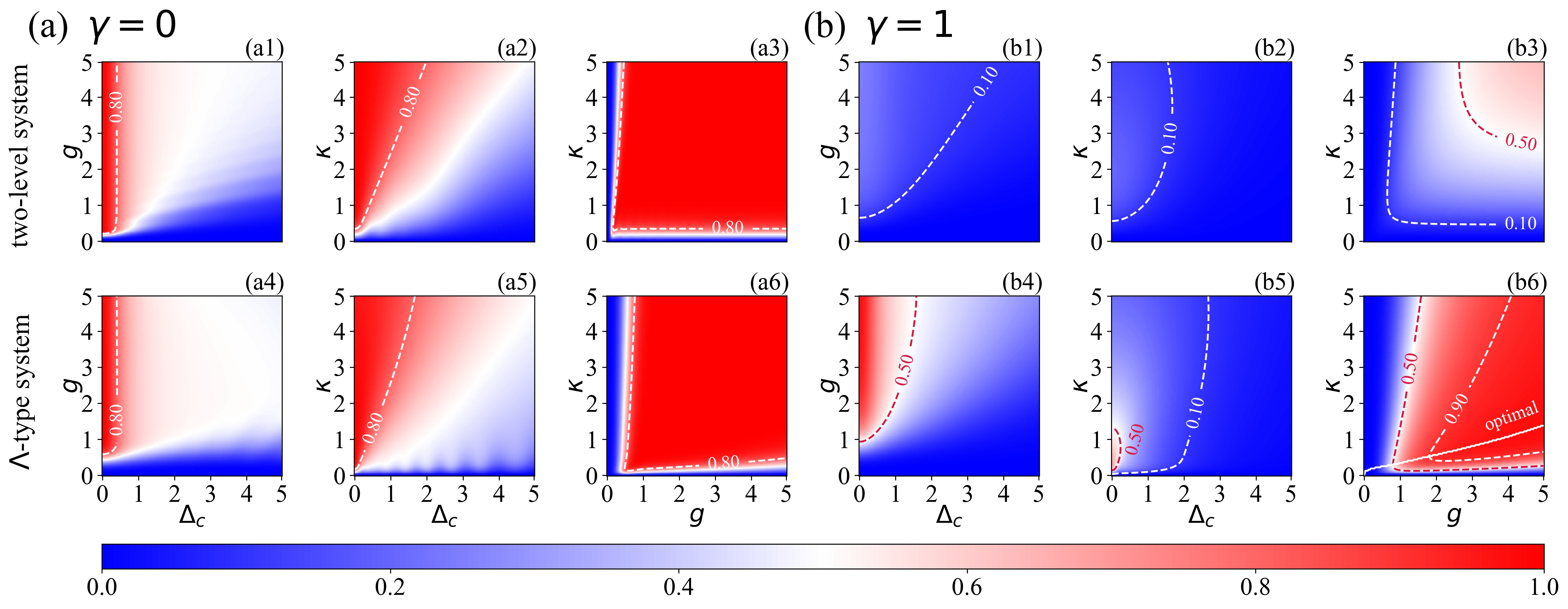} \\
  \caption{HOM interference visibility $V$ dependence on various system parameters for identical CQED-based single-photon sources in the (a) absence and (b) presence of radioactive decay $\gamma$. Dashed lines with labels represent contours indicating corresponding $V$ values, and the solid line in (b6) represents the parameter combinations corresponding to the maximum interference visibility.}
  \label{fig:FIG4}
  \end{figure*}

For both two-level and $\Lambda$-type CQED systems, the emitted single-photon properties are mainly affected by four system parameters: $\Delta_{\text{c}}$, $\kappa$, $g$ and $\gamma$. It is crucial to understand the influence of these parameters on the indistinguishability of photons. To do so, we calculate the HOM interference visibility versus various system parameters for both two-level and $\Lambda$-type CQED systems. In the following simulation, we consider the system parameters normalized to unity with respect to a fixed frequency.

We first investigate HOM interference visibility $V$ between identical CQED systems. In Fig.~\ref{fig:FIG4}, we show the dependence of $V$ on $\{\Delta_{\text{c}},\kappa,g\}$ in the absence and presence of radioactive decay $\gamma$, here we take $\gamma=1$ for the latter case. When we perform the sweep of different system parameters, the other parameter takes constant values: $\Delta_{\text{c}} = 0$, $g=1$ and $\kappa=1$. In the case of the $\Lambda$-type CQED system, we apply a driving field with a Gaussian-type form $\Omega_{\text{d}}(t) = 6\exp{-(\frac{t-15}{5})^2}$ and consider its frequency detuning to be zero ($\Delta_{\text{d}}=0$).

As can be seen from Fig.~\ref{fig:FIG4}(a), in the absence of radioactive decay ($\gamma = 0$), the dependence of $V$ on different system parameters in the two-level CQED system is similar to that in the $\Lambda$-type CQED system. For both types of CQED systems, the value of visibility $V$ decreases rapidly with the detuning $\Delta_{\text{c}}$, as shown in Fig.~\ref{fig:FIG4}(a1, a2, a4 and a5). This indicates that the indistinguishability of single photons generated by the CQED system is highly sensitive to the cavity-atom detuning even when there is no radioactive decay. The CQED system using artificial atoms suffers from the inherent randomness in the self-assembled growth process. This process results in a unique structure for each artificial atom. This nonidentical atoms exhibit resonance frequency fluctuations in artificial atoms and subsequently lead to degradation on indistinguishability of the emitted single-photons according to our simulations. As an alternative, the CQED system using natural atoms can ensure identical resonance frequency because all atoms intrinsically have the same properties, thus showcasing the superiority of indistinguishability over the CQED system using artificial atoms.

Moreover, we find that the detuning range enabling high HOM interference visibility increases with $\kappa$, while increasing $g$ has no significant effect on this range. This suggests that the degradation of photon indistinguishability caused by detuning can be mitigated by increasing the cavity decay rate $\kappa$. Under resonance conditions, the HOM interference visibility sharply approaches unity as both the parameters $g$ and $\kappa$ increase, as shown in Fig.~\ref{fig:FIG4}(a3,a6). Compared to the $\Lambda$-type CQED system, the parameter region of $\{g,\kappa\}$ where the two-level CQED system can achieve high values of $V$ is larger. This indicates that two-level CQED-based single-photon sources with identical properties have less requirements on cavity properties to achieve high single-photon indistinguishability in the absence of radioactive decay.

On the other hand, in the presence of radioactive decay ($\gamma=1$), HOM interference visibility between two-level CQED systems is significantly reduced over a wide range of system parameter (Fig.~\ref{fig:FIG4}(b1-b3)), while $\Lambda$-type CQED systems maintain relatively high HOM interference visibility (Fig.~\ref{fig:FIG4}(b4-b6)). From the contours of $V=0.5$ and $V=0.9$ in Fig.~\ref{fig:FIG4}(b6) we can observe that when $g$ is fixed, the HOM interference visibility initially increases and then decreases as $\kappa$ increases. This indicates the existence of an optimal combination of $g$ and $\kappa$ that leads to maximum single-photon indistinguishability. The optimal combination of $g$ and $\kappa$ is represented by a solid line in Fig.~\ref{fig:FIG4}(b6). Since the dynamic control of cavity decay and cavity-atom coupling strength in CQED systems remains a significant challenge, finding the optimal combination of $\{g,\kappa\}$ can provide valuable guidance for engineering reliable CQED-based single-photon sources.

\begin{figure*}
  \centering
  \includegraphics[width=1\linewidth]{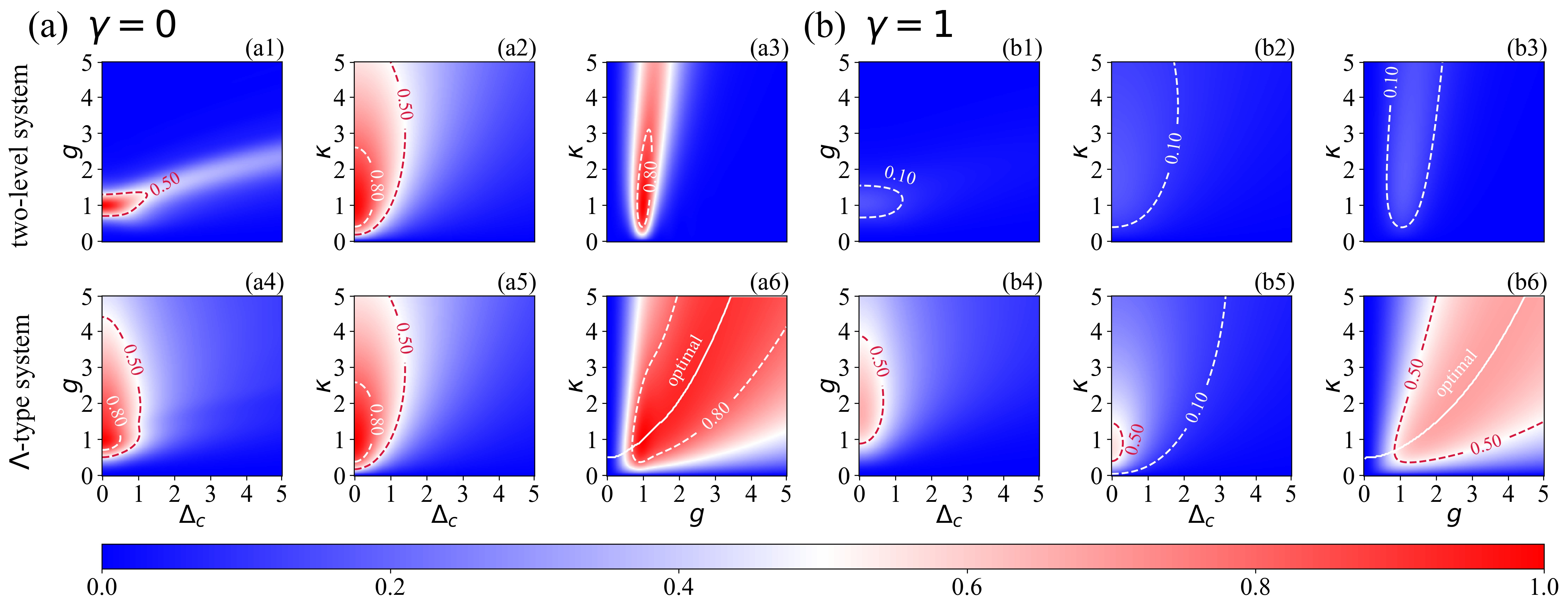} \\
  \caption{HOM interference visibility $V$ dependence on various system parameters for different CQED-based single-photon sources in the (a) absence and (b) presence of radioactive decay $\gamma$.Dashed lines with labels indicate contours representing corresponding $V$ values, and solid lines represent the parameter combinations corresponding to the maximum interference visibility.}
  \label{fig:FIG5}
  \end{figure*}  

In realistic applications, it is challenging to ensure that multiple CQED systems are exactly identical. Hence, it is practically significant and necessary to investigate the indistinguishability of photons between CQED-based single-photon sources with non-identical system parameters. To achieve this, we fix one CQED system in HOM interference as the reference system and then simulate the interference visibility dependence on the parameters of the other CQED system (referred to the interfered system). We choose the following parameters for the reference two-level and $\Lambda$-type CQED systems: $\{\kappa = 1, g = 1, \Delta_{\text{c}}=0\}$.

Numerically simulated HOM interference visibility without radioactive decay is presented in Fig.~\ref{fig:FIG5}(a), from which we can see that $V$ reaches unity when the interfered system parameters are the same as the reference system. This also shows the superiority of natural atoms because they can maintain the identical atomic property. However, there are significant differences in the dependence of $V$ between two-level and $\Lambda$-type CQED systems. For the two-level CQED system, the value of $V$ decreases rapidly as the difference in $g$ between the interfered and reference systems increases, while maintaining a high value of $V$ for $\kappa$ with a larger difference, as shown in Fig.~\ref{fig:FIG5}(a1,a2). On the other hand, the $\Lambda$-type CQED system shows better robustness to differences in $g$ and $\kappa$ (Fig.~\ref{fig:FIG5}(a4,a5)). This can be clearly seen from the dependence of $V$ on $g$ and $\kappa$ under resonance conditions, under which the two-level CQED system can only achieve high HOM interference visibility with system parameters close to the reference system, while a $\Lambda$-type CQED system can maintain high visibility under parameter conditions that are significantly different from the reference system (Fig.~\ref{fig:FIG5}(a3,a6)).

In the presence of radioactive decay ($\gamma=1$), $V$ of both types of CQED systems decreases throughout the entire parameter space. However, the reduction of $V$ in the $\Lambda$-type CQED system is relatively smaller than in the two-level CQED system, as shown in Fig.~\ref{fig:FIG5}(b). Furthermore, the patterns of $V$'s dependence on various parameters are similar to those in the absence of radioactive decay. We demonstrate that the curves for the $\{g,\kappa\}$ combination corresponding to optimal $V$ values with the resonance condition exhibit similar trends in both cases, with and without considering radioactive decay. This suggests that radioactive decay has minimal impact on the optimal combination of system parameters. Therefore, when two CQED-based single photon sources are non-identical, the system with the optimal parameter combination demonstrates greater resilience to radioactive decay.

\section{ML-enhanced single-photon sources}\label{sec:MLresult}

To demonstrate the effectiveness of our ML framework in realistic situations, we evaluate its performance in the $\Lambda$-type CQED system with radioactive decay rate of $\gamma = 1$. We fix one CQED system in the HOM interference scheme unchanged as the reference system and utilize our ML framework to optimize the driving field of the other CQED system with different system parameters. Here, we choose the reference system parameters from the previously mentioned optimal combination: $\{\kappa = 1.25, g = 5, \Delta_{\text{c}}=0, \Delta_{\text{d}}=0\}$. The driving field for the reference system is the same as in Sec.~\ref{sec:Analysis}.

\begin{figure}
  \centering
  \includegraphics[width=1\linewidth]{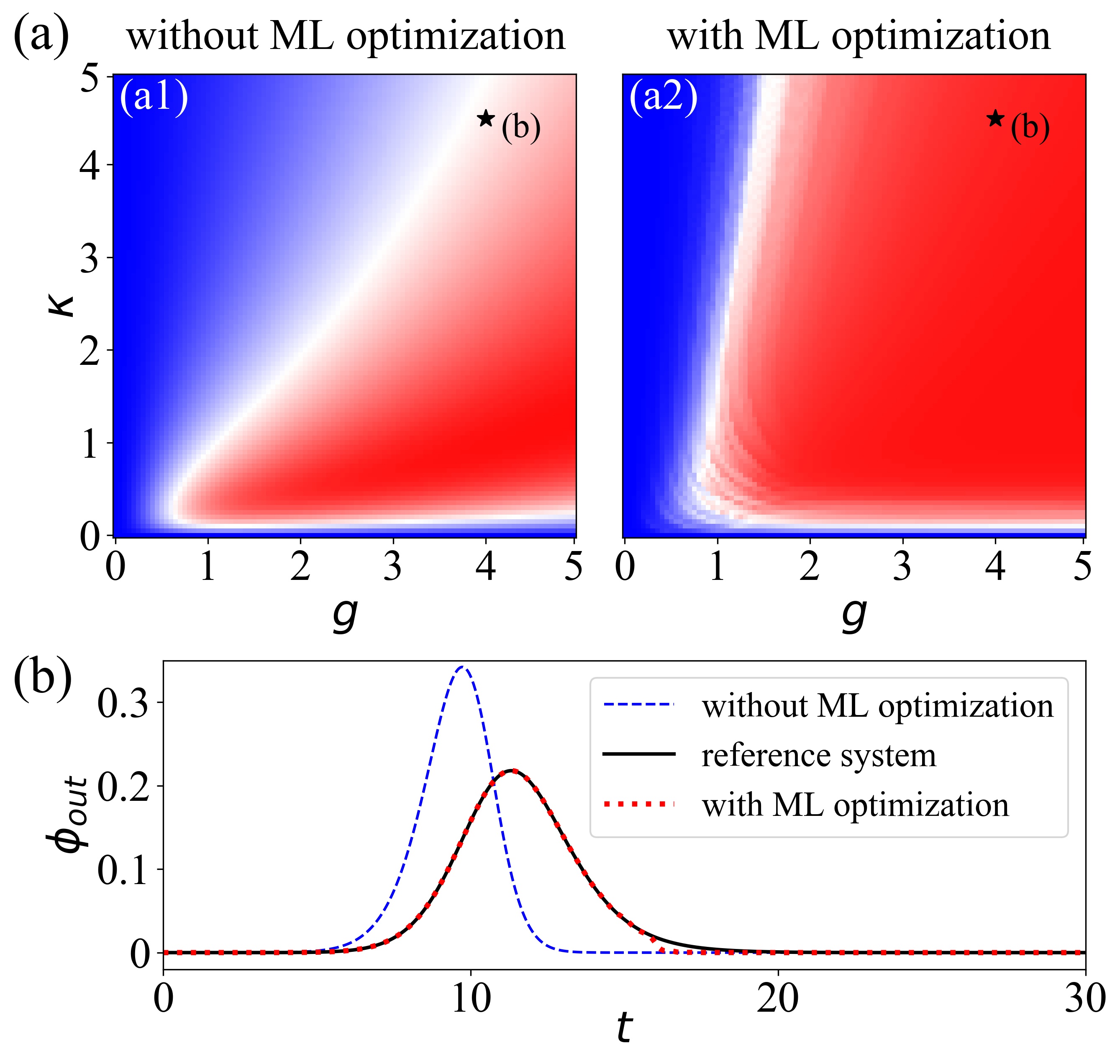} \\
  \caption{Enhancing single-photon indistinguishability with our ML framework. (a) Interference visibility dependence on various CQED system parameters before and after applying ML optimization. (b) The wavefunctions of the emitted single-photon triggered by driving field with and without ML optimization, comparing to that from the reference system. }
  \label{fig:FIG6}
  \end{figure} 

Among all parameters of the CQED system, the strength of atom-cavity coupling and cavity decay are particularly difficult to accurately regulate. Therefore, we apply our ML algorithm to the $\Lambda$-type CQED system with varying values of $g$ and $\kappa$ to determine the optimal driving field that maximizes the HOM interference visibility, thereby enhancing single-photon indistinguishability. The ranges of variation for $g$ and $\kappa$ are $g\in \left(0,5\right]$ and $\kappa\in \left(0,5\right]$, respectively.

The HOM interference visibility with and without ML optimization are shown in Fig.~\ref{fig:FIG6}(a). By comparing Fig.~\ref{fig:FIG6}(a1) and (a2), we can see that our ML optimization has nearly doubled the parameter area for $\{g, \kappa\}$ with high $V$ values. In order to intuitively demonstrate the effectiveness of our ML framework for optimization, we present detailed optimization results for a specific CQED system with parameters $\{g=4,\kappa=4.5\}$ in Fig.~\ref{fig:FIG6}(b). Obviously, the emitted single-photon wavefunction without ML optimization is far different from the wavefunction generated by the reference system. In contrast, the ML-optimized driving field triggers the single-photon wavefunction, which nearly completely overlaps with the wavefunction emitted from the reference system. The $\{g,\kappa\}$ coordinates corresponding to this CQED system are also indicated by a black open circle in Figs.~\ref{fig:FIG6}(a1) and (a2), we can observe that our ML framework substantially enhances the value of $V$ from approximately $0.5$ to over $0.9$.

The aforementioned results indicate that our ML optimization framework can effectively enhance the indistinguishability of single photons generated by CQED systems across a wide range of parameters. In addition, the optimization proposal obtained using our framework can be implemented by adjusting the driving fields without making any changes to the existing CQED systems. This demonstrates the high flexibility and versatility of our ML framework in real-world applications.

\section{Conclusions}\label{sec:discussion}
In conclusion, we have investigated the influence of different CQED system parameters on the indistinguishability of single photons by simulating HOM interference between two independent CQED-based single-photon sources. We have clearly demonstrated that a CQED system using natural atoms has advantages of single-photon indistinguishability over its counterpart using artificial atoms. We have also shown that, for a $\Lambda$-type CQED system, the appropriate selection of detuning, cavity decay, and atom-cavity coupling strength can improve photon indistinguishability and enhance robustness against variations in system parameters and imperfections in system fabrication. Furthermore, we have proposed a ML framework for identifying the optimal driving field to optimize the photon indistinguishability of CQED system. With ML-optimized driving fields, non-identical CQED systems can also generate nearly identical single photons without altering system parameters. Our work opens up a new avenue to engineer scalable and reliable single-photon sources, which may facilitate the implementation of scalable photon-based quantum technologies.

\section*{Acknowledgements}
This work was supported by the National Key R\&D Program of China (Grant No. 2019YFA0308700), the National Natural Science Foundation of China (Grants No. 92365107 and No.11890704), the Program for Innovative Talents and Teams in Jiangsu (Grant No. JSSCTD202138).

%
%%%%%%%%%%%%%%%%%%%%%%%%%%%%%%%%%%%%%%%%%%%%%%%%%%%%%%%%%
%\bibliographystyle{apsrev4-2}
%\bibliography{Reference.bib}

\providecommand{\noopsort}[1]{}\providecommand{\singleletter}[1]{#1}%

\end{document}